\documentclass[preprint]{revtex4}

\usepackage{amssymb,amsmath,bm}
\usepackage{multirow}
\usepackage{color}
\usepackage{graphicx}
\usepackage{cancel}

\begin{document}

\title{Win-Stay-Lose-Shift as a self-confirming equilibrium in the iterated
Prisoner's Dilemma}

\author{
Minjae~Kim$^{1}$, Jung-Kyoo~Choi$^{2}$ and Seung~Ki~Baek$^{1}$}

\address{$^{1}$Department of Physics, Pukyong National University, Busan 48513,
Korea\\
$^{2}$Department of Economics, Kyungpook National University,
Daegu 41566, Korea}

%
%

\begin{abstract}
Evolutionary game theory assumes that players replicate a highly scored player's strategy through genetic inheritance. However, when learning occurs culturally, it is often difficult to recognize someone's strategy just by observing the behaviour. In this work, we consider players with memory-one stochastic strategies in the iterated prisoner's dilemma, with an assumption that they cannot directly access each other's strategy but only observe the actual moves for a certain number of rounds. Based on the observation, the observer has to infer the resident strategy in a Bayesian way and chooses his or her own strategy accordingly. By examining the best-response relations, we argue that players can escape from full defection into a cooperative equilibrium supported by Win-Stay-Lose-Shift in a self-confirming manner, provided that the cost of cooperation is low and the observational learning supplies sufficiently large uncertainty.
\end{abstract}


\maketitle

\section{Introduction}
Evolutionary game theorists often assume that
behavioural traits
can be genetically transmitted across generations~\cite{smith1982evolution}.
Along this line, researchers have investigated the genetic basis of
cooperative behaviour~\cite{kasper2017genetics,manfredini2018candidate}.
However, humans learn many culture-specific behavioural rules through
observational learning~\cite{bandura1977social}, and this mechanism
mediates ``cultural'' transmission
that has been proved to exist among a number of non-human animals as
well~\cite{krutzen2005cultural,frith2012mechanisms}. The mirror neuron
research suggests that the primate brain may even have a
specialized circuit for imitating each other's behaviour, which facilitates social
learning~\cite{di1992understanding,gallese1996action,ferrari2014mirror}.
%
%
%
In comparison with the direct genetic transmission, the non-genetic inheritance
through social learning can provide better adaptability by responding faster to
environmental changes~\cite{leimar2015evolution}.

In contrast with genetic inheritance, however, observational learning may lead
to imperfect mimicry if observation is not sufficiently informative or
involved with a systematic bias. The notion of self-confirming equilibrium (SCE)
has been proposed by incorporating such imperfectness of observation in
learning~\cite{fudenberg1998theory}: When a SCE strategy is played, some of
the possible information sets may not be reached, so the players do not have
exact knowledge but only certain untested belief about what their co-players
would do at those unreached sets. It is nevertheless sustained as an equilibrium
in the sense that no player can expect a better payoff by unilaterally deviating
from it once given such belief, and that the beliefs do not conflict with
observed moves.
Dynamics of learning based on a limited set of information has been
investigated in the context of the coordination
game~\cite{sandholm2001almost,kreindler2013fast}, in which
the opponent's observed decision is assumed to be his or her strategy.
However,
the subtlety of cultural transmission manifests itself clearly when a strategy
is regarded as a decision rule, hidden from the observer, rather than
the decision itself.

In this work, we investigate the iterated prisoner's dilemma (PD) game among
players with memory-one strategies, who infer the resident strategy from
observation and optimizes their own strategies against it. By memory-one, we
mean that a player refers to the previous round to choose a move between
cooperation and defection~\cite{baek2016comparing}.
If we restrict ourselves to memory-one strategies, it is already
well known in evolutionary game theory that `Win-Stay-Lose-Shift
(WSLS)'~\cite{kraines1989pavlov,nowak1993strategy,imhof2007tit} can appear
through mutation and take over the population from defectors if the cost of
cooperation is low~\cite{baek2016comparing}. Compared with such an evolutionary
approach, we will impose ``less bounded'' rationality in that our players are
assumed to be capable of computing the best response to a given strategy within
the memory-one pure-strategy space.
We will identify the best-response dynamics in this space
and examine how the dynamics should be modified when observational learning
introduces uncertainty in Bayesian inference about strategies. If every player
exactly replicated each other's strategy, full defection would be a Nash
equilibrium (NE) for any cost of cooperation. Under uncertainty in observation,
however, our finding is that defection is not always a SCE so that the
population can move to a cooperative equilibrium supported by WSLS, which is
both a SCE and a NE and can thus be called a SCENE.

\section{Method and Result}

\subsection{Best-response relations without observational uncertainty}

Let us define the one-shot PD game in the following form:
\begin{equation}
\left(
\begin{array}{c|cc}
   & C & D\\\hline
C  & 1-c & -c\\
D  & 1 & 0
\end{array}
\right),
\label{eq:payoff}
\end{equation}
where we abbreviate cooperation and defection as $C$ and $D$, respectively, and
$c$ is the cost of cooperation assumed to be $0<c<1$. In this work, the game of
Eq.~\eqref{eq:payoff} will be repeated indefinitely. Furthermore, the
environment is noisy: Even if a player intends to cooperate, it
can be misimplemented as defection, or vice versa, with probability $\epsilon$.
In the analysis below, we will take $\epsilon$ as an arbitrarily small positive
number.

We will restrict ourselves to the space of memory-one (M$_1$) pure strategies.
By a M$_1$ pure strategy, we mean that it chooses a move between $C$ and $D$
as a function of the two players' moves in the previous round. We thus describe
such a strategy as $[p_{_{CC}}, p_{_{CD}}, p_{_{DC}}, p_{_{DD}}]$, where
$p_{_{XY}}=1$ means that $C$ is prescribed when the players
did $X$ and $Y$, respectively, in the previous round, and
$p_{_{XY}}=0$ if $D$ is prescribed in the same situation.
Note that the initial move in the first round is irrelevant to the long-term
average payoff in the presence of error so that it has been discarded in the
description of a strategy. The set of M$_1$ pure strategies, denoted by
$\Delta$, contains $16$ elements from $\mathbf{d}_0 \equiv [0,0,0,0]$
to $\mathbf{d}_{15}\equiv [1,1,1,1]$.

Let us assume that a player, say, Alice, takes a M$_1$ pure strategy
$\mathbf{d}_\alpha$ as her strategy.
The noisy environment
effectively modifies her behaviour to
\begin{equation}
\mathbf{s}_A^\epsilon \equiv
(1-\epsilon) \mathbf{d}_\alpha + \epsilon (\mathbf{1}-\mathbf{d}_\alpha)
\label{eq:effective}
\end{equation}
as if she were playing a mixed strategy,
where $\mathbf{1} \equiv [1,1,1,1]$. Likewise, Alice's co-player Bob
chooses $\mathbf{d}_\beta$, and his effective behaviour is described by
\begin{equation}
\mathbf{s}_B^\epsilon \equiv (1-\epsilon)
\mathbf{d}_\beta + \epsilon (\mathbf{1}-\mathbf{d}_\beta).
\label{eq:effective2}
\end{equation}
The repeated interaction between Alice and Bob is Markovian, and it is
straightforward to obtain the stationary probability distribution
\begin{equation}
\mathbf{v} (\mathbf{d}_\alpha, \mathbf{d}_\beta, \epsilon) = (v_{_{CC}},
v_{_{CD}}, v_{_{DC}}, v_{_{DD}}),
\label{eq:stationary}
\end{equation}
where $v_{_{XY}}$ means the long-term
average probability to observe Alice and Bob choosing $X$ and $Y$,
respectively~\cite{nowak1990stochastic,nowak1995automata,press2012iterated}
(see Appendix~\ref{app:stationary} for more details).
The presence of $\epsilon>0$ guarantees the uniqueness of $\mathbf{v}$.
Alice's long-term average payoff against Bob is then calculated as
\begin{equation}
\Pi(\mathbf{d}_\alpha, \mathbf{d}_\beta, \epsilon) = \mathbf{v} \cdot\mathbf{P}
\label{eq:avgpay}
\end{equation}
where $\mathbf{P} \equiv (1-c, -c, 1, 0)$ is a payoff vector corresponding to
Eq.~\eqref{eq:payoff}. As long as Alice can exactly identify Bob's strategy
$\mathbf{d}_\beta$ with no observational uncertainty, she can find the best
response to Bob within the set of M$_1$ pure strategies by applying every
$\mathbf{d}_\alpha \in \Delta$ to Eq.~\eqref{eq:avgpay}.


\begin{table}
\begin{tabular}{cccc}\\\hline
Opponent & Best & Payoff of the best response
& \multirow{2}{*}{Misc.}\\
strategy & response & to the opponent strategy & \\\hline
$\mathbf{d}_{0}$ & $\mathbf{d}_{0}^\dagger$  & $(1-c)\epsilon$ & AllD\\
$\mathbf{d}_{1}$ & $\mathbf{d}_{0}$  & $1/2 - (1/4+c)\epsilon + O(\epsilon^2)$ & \\
$\mathbf{d}_{2}$ & $\mathbf{d}_{11}$ & $(1-c)/2-(1+c)\epsilon/2 + O(\epsilon^2)$ & \\
$\mathbf{d}_{3}$ & $\mathbf{d}_{0}$  & $1/2-ce + O(\epsilon^3)$ & \\
$\mathbf{d}_{4}$ & $\mathbf{d}_{0}$  & $1/3+\left(2/9 -c\right)\epsilon + O(\epsilon^2)$ & \\
$\mathbf{d}_{5}$ & $\mathbf{d}_{0}$  & $1-(2+c)\epsilon + O(\epsilon^2)$ & \\
$\mathbf{d}_{6}$ & $\mathbf{d}_{9}$  & $1-3(1+c)\epsilon + O(\epsilon^2)$ & \\
$\mathbf{d}_{7}$ & $\mathbf{d}_{0}$  & $1-(2+c)\epsilon+4\epsilon^2 + O(\epsilon^3)$ & \\
$\mathbf{d}_{8}$ & $\left\{\begin{array}{ll} \mathbf{d}_{8}^\dagger, & c>1/3 \\
\mathbf{d}_{15}, & c< 1/3 \end{array}\right\}$  &
 $\left\{\begin{array}{c}
 3(1-c)\epsilon/2 + O(\epsilon^2)\\
 1/3-c + O(\epsilon)
 \end{array}\right\}$ & GT$_1$\\
$\mathbf{d}_{9}$ & $\left\{\begin{array}{ll} \mathbf{d}_{0}, & c>1/2 \\
\mathbf{d}_{9}^\dagger, & c< 1/2 \end{array}\right\}$ &
 $\left\{\begin{array}{c}
 1/2 + O(\epsilon)\\
 1-c + O(\epsilon)
 \end{array}\right\}$ & WSLS\\
 $\mathbf{d}_{10}$ & $\mathbf{d}_{15}$  & $(1-c)-(2-c)\epsilon + O(\epsilon^2)$ &
 TFT\\
 $\mathbf{d}_{11}$ & $\left\{\begin{array}{ll} \mathbf{d}_{0}, & c>1/2 \\
 \mathbf{d}_{13}, & c< 1/2 \end{array}\right\}$ &
$\left\{\begin{array}{c}
 1/2+\left(1/4-c\right)\epsilon + O(\epsilon^2)\\
 (1-c)-(2-c)\epsilon + O(\epsilon^2)
 \end{array}\right\}$ & \\
 $\mathbf{d}_{12}$ & $\mathbf{d}_{0}$  & $1/2 + O(\epsilon)$ & \\
 $\mathbf{d}_{13}$ & $\mathbf{d}_{0}$  & $1-(1+c)\epsilon + O(\epsilon^2)$ & \\
 $\mathbf{d}_{14}$ & $\mathbf{d}_{1}$  & $1-2(1+c)\epsilon + O(\epsilon^2)$ & \\
 $\mathbf{d}_{15}$ & $\mathbf{d}_{0}$  & $1-(1+c)\epsilon + O(\epsilon^3)$ & AllC\\\hline
\end{tabular}
\caption{Best response among M$_1$ pure strategies. Against each
strategy in the first column, we obtain the best response (the second column),
and the resulting average payoff [Eq.~\eqref{eq:avgpay}] earned
by the best response is given as a power series of $\epsilon$ in the third
column. In the second column, we have placed a dagger next to a strategy when it
is the best response to itself.}
\label{table:br}
\end{table}

\begin{figure}
\includegraphics[width=\textwidth]{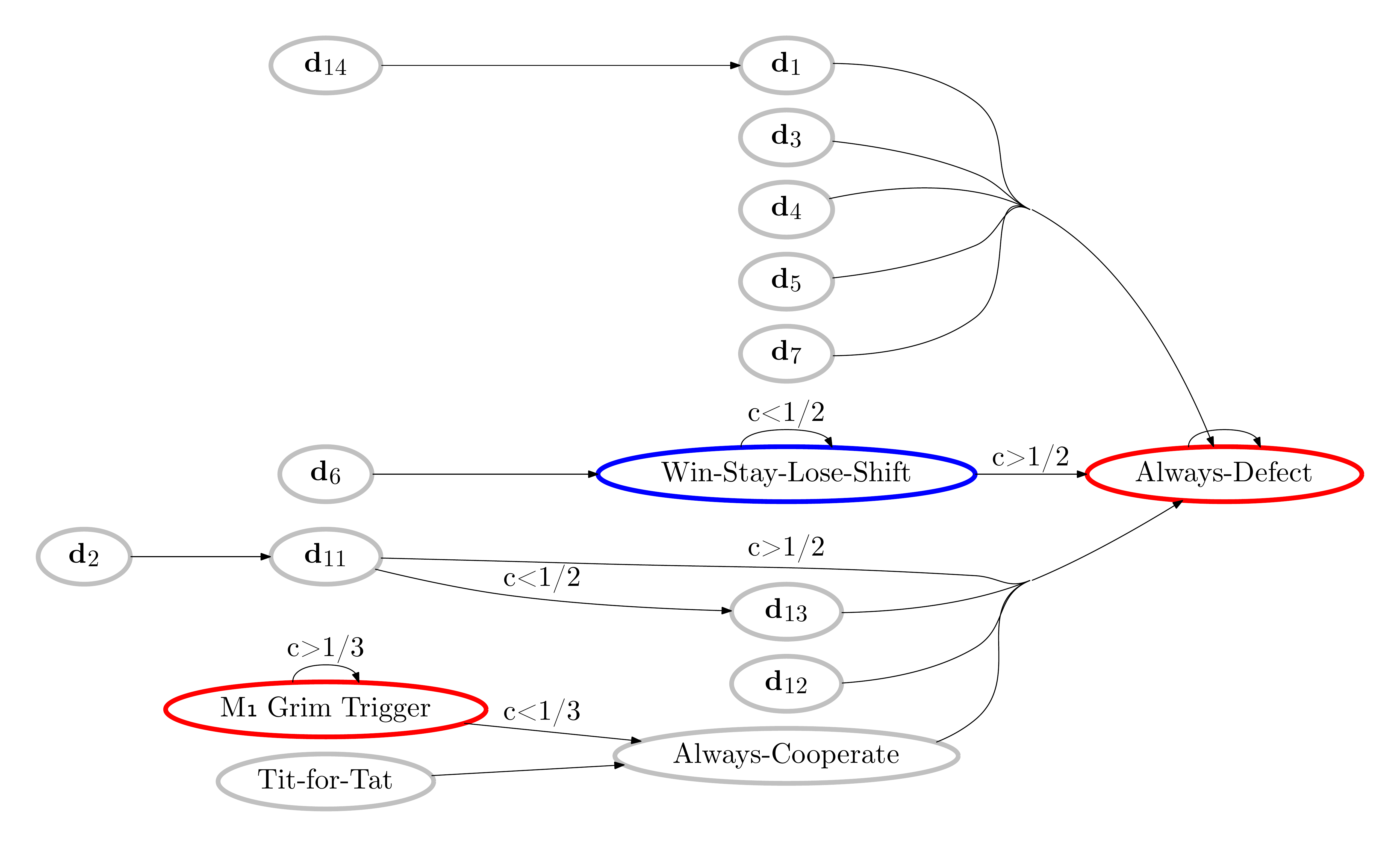}
\caption{Graphical representation of best-response relations in
Table~\ref{table:br}. If $\mathbf{d}_\mu$ is
the best response to $\mathbf{d}_\nu$, we represent it as an arrow
from $\mathbf{d}_\nu$ to $\mathbf{d}_\mu$. The blue node means an efficient NE
with $1-v_{_{CC}} \sim O(\epsilon)$, whereas the red nodes mean
inefficient ones with $v_{_{CC}} \lesssim O(\epsilon)$ as shown in
Table~\ref{table:vXY}.}
\label{fig:best16}
\end{figure}

In Table~\ref{table:br}, we list the best response to each strategy in $\Delta$
in the limit of small $\epsilon$ (see also Fig.~\ref{fig:best16} for its
graphical representation). In most cases, the best-response dynamics
ends up with $\mathbf{d}_0 = [0,0,0,0]$, which is the best response to
itself and often called Always-Defect (AllD). For example, if we start with
Tit-for-Tat (TFT), represented as $\mathbf{d}_{10} = [1,0,1,0]$,
Table~\ref{table:br} shows that the best response to TFT
within $\Delta$ is Always-Cooperate (AllC), represented as $\mathbf{d}_{15} =
[1,1,1,1]$, to which AllD is the best response for obvious reasons.

However, two exceptions exist: The first one is $\mathbf{d}_8 =
[1,0,0,0]$, which we may call M$_1$ Grim Trigger (GT$_1$). If $c > 1/3$, this
strategy is the best response to itself, and it is an inefficient
equilibrium giving each player an average payoff of $O(\epsilon)$.
The other exception is WSLS, represented by
$\mathbf{d}_{9} = [1,0,0,1]$, which is the best response to itself when $c \le
1/2$. It is an efficient NE, at which each player earns $1-c + O(\epsilon)$ per
round on average.

\subsection{Observational learning}

Now, let us imagine a monomorphic population of players who have adopted a
strategy $\mathbf{d}_\gamma$ in common. The population is in equilibrium in the
sense that a large ensemble of their states $XY \in \{CC, CD, DC, DD\}$ can
represent the stationary probability distribution $\mathbf{v} (\mathbf{d}_\gamma,
\mathbf{d}_\gamma, \epsilon)$.
We have an observer, say, Alice, with a potential strategy $\mathbf{d}_\alpha$. As
we learn social norms in childhood, it is assumed that Alice does not yet
participate in the game but has a learning period to observe $M (\gg 1)$ pairs
of players, all of whom have used the resident strategy $\mathbf{d}_\gamma$.
How their mind works is a black box to her:
Just by observing their states $XY$ and subsequent moves,
Alice has to form belief about $\mathbf{d}_\gamma$, based on which she chooses
her own strategy $\mathbf{d}_\alpha$ to maximize the expected payoff.
If Alice's optimal strategy turns out to be identical to the resident strategy
$\mathbf{d}_\gamma$, it constitutes a SCE.

\begin{table}
\begin{tabular*}{\textwidth}{c|c|@{\extracolsep{\fill}}cccc}\hline
Category & Strategy & $v_{_{CC}}$ & $v_{_{CD}}$ & $v_{_{DC}}$ & $v_{_{DD}}$\\\hline\hline
\multirow{8}{*}{I}
& $\mathbf{d}_{3} = [\bm{\mathsf{0}},\bm{\mathsf{0}},\bm{\mathsf{1}},\bm{\mathsf{1}}]$ & \multirow{4}{*}{$\frac{1}{4}$} & \multirow{4}{*}{$\frac{1}{4}$} & \multirow{4}{*}{$\frac{1}{4}$} & \multirow{4}{*}{$\frac{1}{4}$}\\
& $\mathbf{d}_{5} = [\bm{\mathsf{0}},\bm{\mathsf{1}},\bm{\mathsf{0}},\bm{\mathsf{1}}]$ &  &  &  &  \\
& $\mathbf{d}_{10} = [\bm{\mathsf{1}},\bm{\mathsf{0}},\bm{\mathsf{1}},\bm{\mathsf{0}}]$ &  &  &  &  \\
& $\mathbf{d}_{12} = [\bm{\mathsf{1}},\bm{\mathsf{1}},\bm{\mathsf{0}},\bm{\mathsf{0}}]$ &  &  &  &  \\
\cline{2-6}
& $\mathbf{d}_{2} = [0,\bm{\mathsf{0}},\bm{\mathsf{1}},\bm{\mathsf{0}}]$ & \multirow{2}{*}{$\frac{1}{2}\epsilon$} & \multirow{2}{*}{$\frac{1}{4}$} & \multirow{2}{*}{$\frac{1}{4}$} & \multirow{2}{*}{$\frac{1}{2}$} \\
& $\mathbf{d}_{4} = [0,\bm{\mathsf{1}},\bm{\mathsf{0}},\bm{\mathsf{0}}]$ &  &  &  &  \\
\cline{2-6}
& $\mathbf{d}_{11} = [\bm{\mathsf{1}},\bm{\mathsf{0}},\bm{\mathsf{1}},1]$ & \multirow{2}{*}{$\frac{1}{2}$} & \multirow{2}{*}{$\frac{1}{4}$} & \multirow{2}{*}{$\frac{1}{4}$} & \multirow{2}{*}{$\frac{1}{2}\epsilon$} \\
& $\mathbf{d}_{13} = [\bm{\mathsf{1}},\bm{\mathsf{1}},\bm{\mathsf{0}},1]$ &  &  &  &  \\
\hline
\multirow{2}{*}{II}
& $\mathbf{d}_{1} = [\bm{\mathsf{0}},0,0,\bm{\mathsf{1}}]$ & \multirow{2}{*}{$\frac{1}{2}$} & \multirow{2}{*}{$\epsilon$} & \multirow{2}{*}{$\epsilon$} & \multirow{2}{*}{$\frac{1}{2}$} \\
& $\mathbf{d}_{7} = [\bm{\mathsf{0}},1,1,\bm{\mathsf{1}}]$ &  &  &  &  \\
\hline
\multirow{6}{*}{III}
& $\mathbf{d}_{0} = [0,0,0,\bm{\mathsf{0}}]$ & $\epsilon^2$ & $\epsilon$ & $\epsilon$ & $1$ \\
& $\mathbf{d}_{6} = [0,1,1,\bm{\mathsf{0}}]$ & $2\epsilon$ & $\epsilon$ & $\epsilon$ & $1$ \\
& $\mathbf{d}_{8} = [1,0,0,\bm{\mathsf{0}}]$ & $\frac{1}{2}\epsilon$ & $\epsilon$ & $\epsilon$ & $1$ \\
\cline{2-6}
& $\mathbf{d}_{9} = [\bm{\mathsf{1}},0,0,1]$ & $1$ & $\epsilon$ & $\epsilon$ & $2\epsilon$ \\
& $\mathbf{d}_{14} = [\bm{\mathsf{1}},1,1,0]$ & $1$ & $\epsilon$ & $\epsilon$ & $\frac{1}{2}\epsilon$ \\
& $\mathbf{d}_{15} = [\bm{\mathsf{1}},1,1,1]$ & $1$ & $\epsilon$ & $\epsilon$ & $\epsilon^2$ \\
\hline
\end{tabular*}
\caption{Stationary probability distribution $\mathbf{v} (\mathbf{d}_\gamma,
\mathbf{d}_\gamma, \epsilon)$, where we have retained only the leading-order term
in the $\epsilon$-expansion for each $v_{_{XY}}$. When we describe a strategy in
binary, the boldface digits are the ones that are frequently observed with
$v_{_{XY}} \sim O(1)$ and thus readily identifiable as long as $M \gg 1$.
In this table, the eight strategies in Category I have three or four such
digits, so if the population is using one of these strategies, Alice can tell
which one is being played after $M (\gg 1)$ observations. As for Category II,
the member strategies $\mathbf{d}_{1}$ and $\mathbf{d}_{7}$ would be
indistinguishable if $M \ll \epsilon^{-1}$ because they differ at their
non-boldface digits. Still, Alice can find the best response $\mathbf{d}_{0}$
which is common to both of them (see Table~\ref{table:br}). In Category III,
each member strategy has just one boldface digit, so the strategies as well as
the best responses can be identified only if $M \gg \epsilon^{-1}$.
}
\label{table:vXY}
\end{table}

To see how Alice can specify $\mathbf{d}_\gamma \in \Delta$ from observation, let
us consider an example that the observed probability distribution over states
$XY$ is best described as $\mathbf{v} \approx (0,1/4,1/4,1/2)$. If Alice has
computed $\mathbf{v}$ for every strategy in $\Delta$ as listed in
Table~\ref{table:vXY}, the observation suggests that the resident strategy is
unlikely to be TFT ($\mathbf{d}_{10} = [1,0,1,0]$) because the corresponding
stationary distribution would be $\mathbf{v} = (1/4, 1/4, 1/4, 1/4)$.
She finds that $\mathbf{d}_\gamma$ can be either $\mathbf{d}_{2} =
[0,0,1,0]$ or $\mathbf{d}_{4} = [0,1,0,0]$. To distinguish between them, she has
to check how people react to $CD$ or $DC$. According to Table~\ref{table:vXY},
these states will be observed frequently because $v_{_{CD}} =
v_{_{DC}} = 1/4$. Thus, in this example, Alice succeeds in identifying
$\mathbf{d}_\gamma$ as long as $M \gg 1$. Eight strategies have this property,
constituting Category I in $\Delta$ (Table~\ref{table:vXY}).
As another example, if $\mathbf{v} \approx (1/2,0,0,1/2)$, Alice sees that
$\mathbf{d}_\gamma$ must be either $\mathbf{d}_{1} = [0,0,0,1]$ or $\mathbf{d}_{7} =
[0,1,1,1]$. To resolve the uncertainty, she has to further
check how people react to $CD$ or $DC$, but she may actually save this effort
because the best response turns out to be $\mathbf{d}_{0}$ in either case
(Table~\ref{table:br}). This is the case of Category II in $\Delta$
(Table~\ref{table:vXY}).

In general, the first important piece of information to infer $\mathbf{d}_\gamma$
is the stationary distribution $\mathbf{v}$ because it heavily depends on
$\mathbf{d}_\gamma$ (Table~\ref{table:vXY}). However, the information of
$\mathbf{v}$ may be insufficient to single out the answer: Suppose that $\mathbf{v}$
gives multiple candidate strategies which prescribe different moves at a
certain state $XY$ and thus have different best responses. Alice then needs to
observe what players actually choose at $XY$, and such observations should be
performed sufficiently many times, i.e., $M v_{_{XY}} \gg 1$, for the sake of
statistical power. If we check every $\mathbf{d}_\gamma \in \Delta$ one by one in
this way, we see that the best response to the resident strategy can readily be
identified as long as $M \gg \epsilon^{-1}$, in which case the result of
observational learning would be the same as that of exact identification of
strategies.

\begin{figure}
\includegraphics[width=\textwidth]{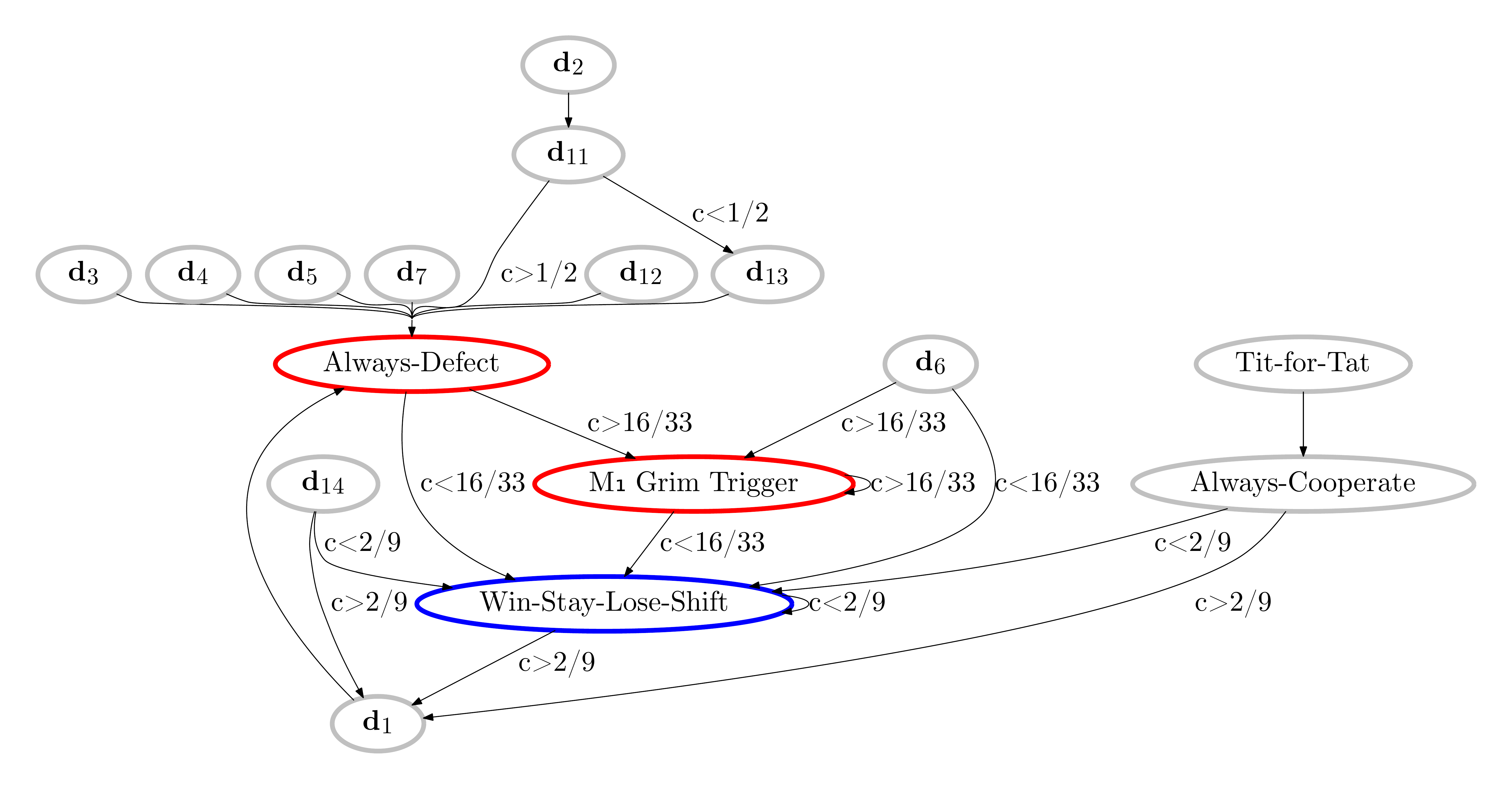}
\caption{Best-looking responses to maximize the expected payoff under
uncertainty in observation, when $1 \ll M
\ll \epsilon^{-1}$. Compared with Fig.~\ref{fig:best16}, the first difference
is that Alice uses Eq.~\eqref{eq:uncertain1} against $\mathbf{d}_{0}$,
$\mathbf{d}_{6}$, and $\mathbf{d}_{8}$. In addition, she will use
Eq.~\eqref{eq:uncertain2} against $\mathbf{d}_{9}$, $\mathbf{d}_{14}$, and
$\mathbf{d}_{15}$.
}
\label{fig:best16_uncertainty}
\end{figure}

If $M \ll \epsilon^{-1}$, on the other hand, Alice cannot fully resolve such
uncertainty through observation. Still, note that $M$ should be taken as far
greater than $O(1)$ for statistical inference to be meaningful.
Furthermore, $\epsilon$ has been introduced as a regularization
parameter whose exact magnitude is irrelevant, so we look at the behaviour
in the limit of small $\epsilon$. When $1 \ll M \ll \epsilon^{-1}$,
uncertainty in the best response remains only when $\mathbf{v} \approx (0,0,0,1)$
or $(1,0,0,0)$, both of which are characteristic of Category III in
Table~\ref{table:vXY}. In the former case, $\mathbf{d}_{0}$, $\mathbf{d}_{6}$, and
$\mathbf{d}_{8}$ are the candidate strategies for $\mathbf{d}_\gamma$ ,
whereas in the latter case, the candidates are $\mathbf{d}_{9}$, $\mathbf{d}_{14}$,
and $\mathbf{d}_{15}$. From the Bayesian perspective, it is reasonable to assign
equal probability to each of the candidate strategies.
However, if $M \epsilon \ll 1$, the number of observations cannot be
enough to update this prior probability (see Appendix~\ref{app:bayes} for a
detailed discussion). Therefore, when $\mathbf{v} \approx (0,0,0,1)$, yielding
$\mathbf{d}_\gamma = \mathbf{d}_{0}$ or $\mathbf{d}_{6}$ or $\mathbf{d}_{8}$,
Alice tries to maximize the expected payoff
\begin{equation}
\overline{\Pi}_\alpha = \frac{\Pi(\mathbf{d}_\alpha, \mathbf{d}_{0}, \epsilon)
+ \Pi(\mathbf{d}_\alpha, \mathbf{d}_{6}, \epsilon)
+ \Pi(\mathbf{d}_\alpha, \mathbf{d}_{8}, \epsilon)}{3},
\label{eq:uniform}
\end{equation}
and the calculation shows that it can be achieved by playing
\begin{equation}
\left\{
\begin{array}{ll}
\mathbf{d}_{8}, & \text{if~}c>16/33 \\
\mathbf{d}_{9}, & \text{if~}c<16/33
\end{array}
\right.
\label{eq:uncertain1}
\end{equation}
in the limit of $\epsilon \to 0$.
Likewise, when $\mathbf{v} \approx (1,0,0,0)$, yielding $\mathbf{d}_\gamma =
\mathbf{d}_{9}$ or $\mathbf{d}_{14}$ or $\mathbf{d}_{15}$, Alice tries to maximize her
expected payoff from the three possibilities, which is achieved when she plays
\begin{equation}
\left\{
\begin{array}{ll}
\mathbf{d}_{1}, & \text{if~}c>2/9\\
\mathbf{d}_{9}, & \text{if~}c<2/9
\end{array}
\right.
\label{eq:uncertain2}
\end{equation}
as $\epsilon \to 0$.
Now, AllD ceases to be the best-looking response to
itself (Fig.~\ref{fig:best16_uncertainty}):
The expected payoff against AllD will be higher when WSLS is played, if
$c<16/33$.
On the other hand, if we consider a WSLS population with $c < 2/9$, its
cooperative equilibrium is protected from invasion of defectors because Alice
under observational uncertainty will keep choosing WSLS, which is truly the best
response to itself.

\begin{figure}
\includegraphics[height=0.36\textwidth,width=0.4\textwidth]{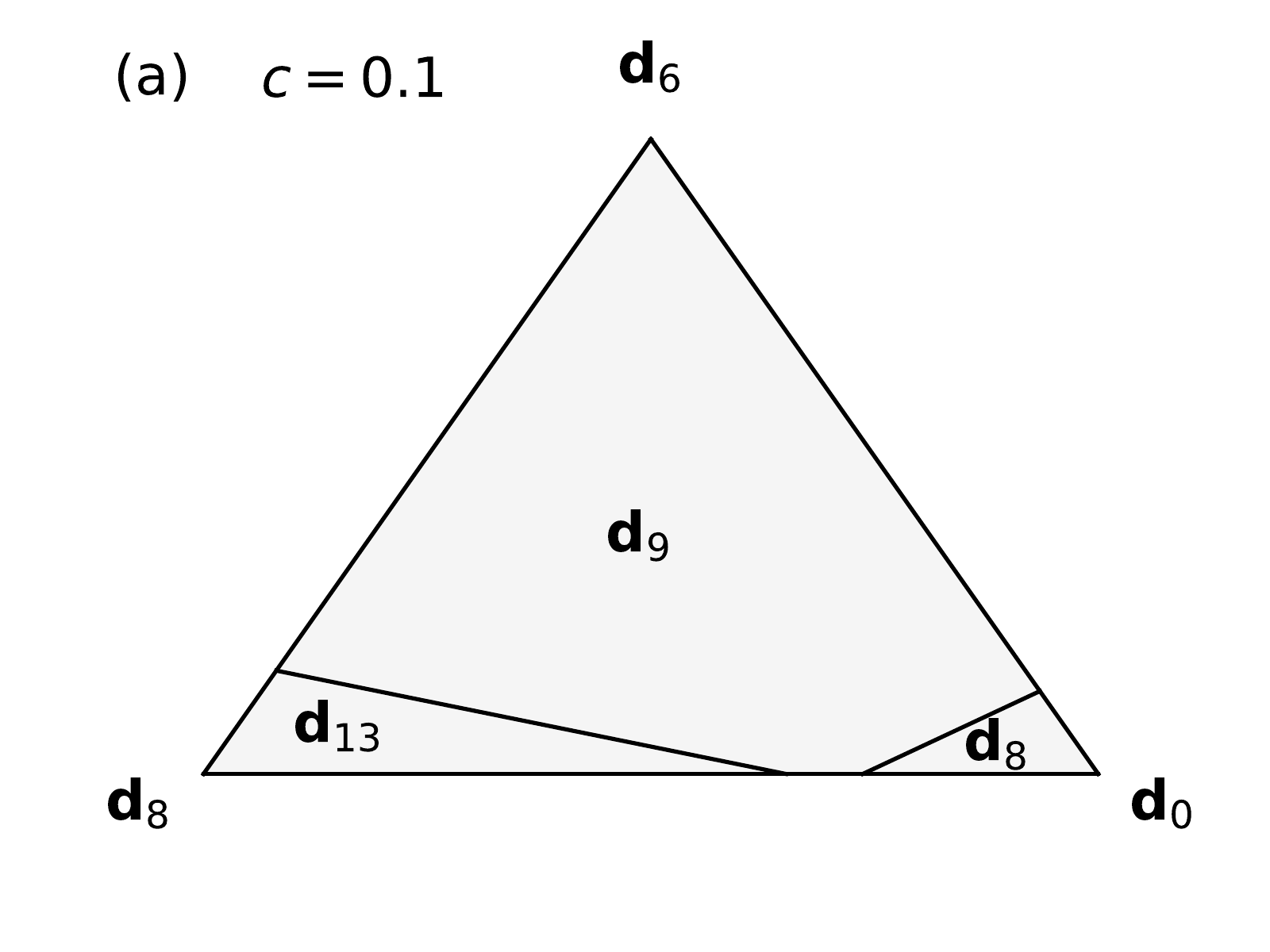}
\includegraphics[height=0.36\textwidth,width=0.4\textwidth]{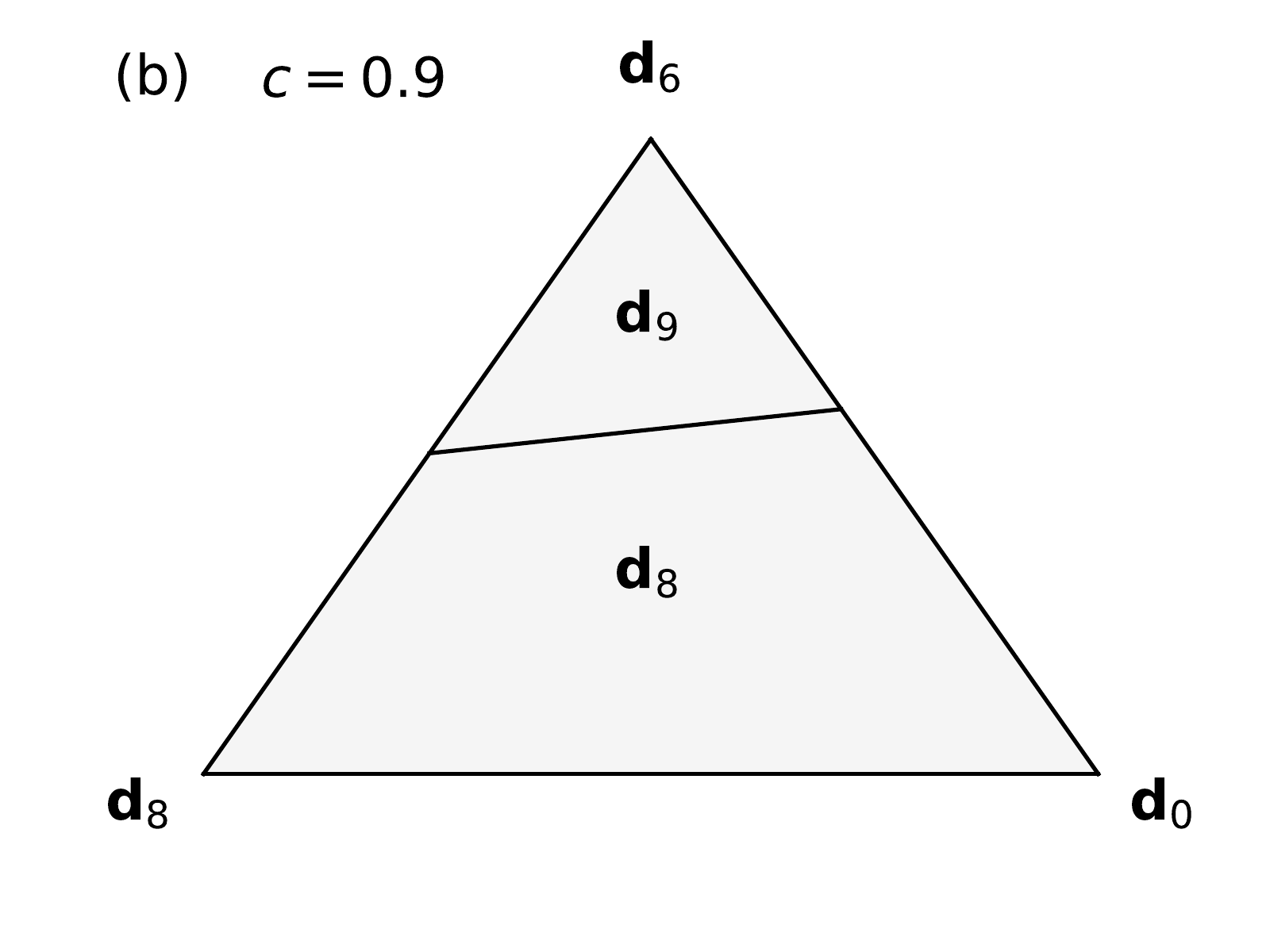}\\
\includegraphics[height=0.36\textwidth,width=0.4\textwidth]{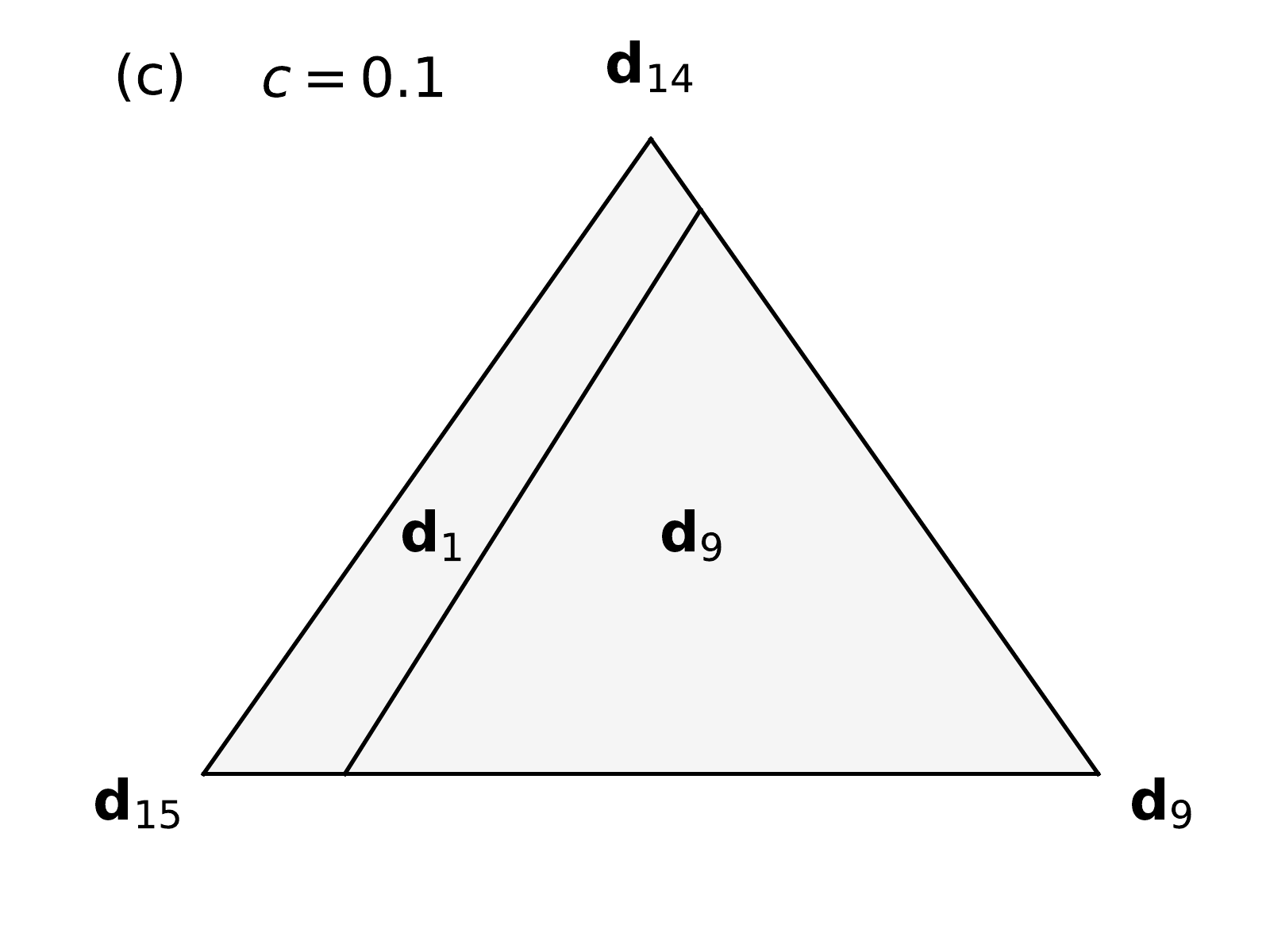}
\includegraphics[height=0.36\textwidth,width=0.4\textwidth]{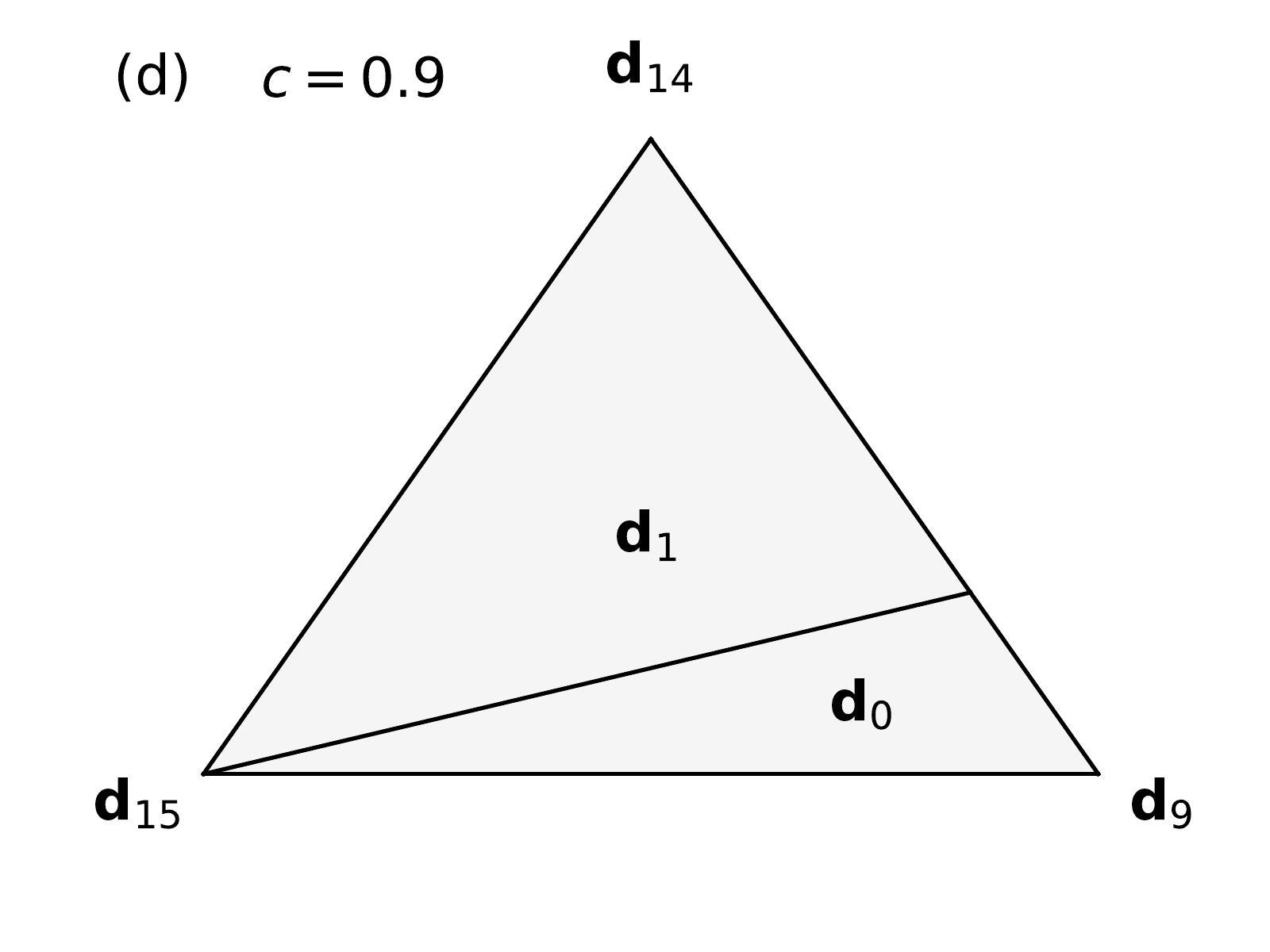}
\caption{Effect of the prior on the observer's choice.
A point in the triangle represents three fractions, which sum up to one, and
its distance to an edge is proportional to the fraction of the strategy at the
opposite vertex~\cite{pythonternary}. (a) When the observer sees nearly defection only, the prior takes
the form of $(f_0, f_6, f_8)$, for which we can find the strategy that gives the
best expected payoff as written in each region.
When $c$ is low, $\mathbf{d}_9$ (WSLS) gives the highest expected payoff for
most of the prior.
(b) Even when the cost increases to $c=0.9$, the observer should choose WSLS
if the prior contains a sufficiently high fraction of
$\mathbf{d}_6$. (c) If the observer sees cooperation almost all the time, the prior can be expressed as
$(f_9, f_{14}, f_{15})$. If $c$ is low, WSLS can be the observer's choice
when $f_9$ is high enough.
(d) The region of WSLS disappears as $c$ exceeds $1/2$, and the only possible
choice is between $\mathbf{d}_1$ and $\mathbf{d}_0$ (AllD).}
\label{fig:simplex}
\end{figure}
The above analysis concerns the uniform prior among three candidate strategies
in each case. Let $f_i$ denote the fraction of $\mathbf{d}_i$. For an observer
who almost always sees defection from the population, the prior in
Eq.~\eqref{eq:uniform} can be written as $(f_0, f_6, f_8) = (1/3, 1/3, 1/3)$.
For a general prior $(f_0, f_6, f_8)$ with $0< f_i <1$ and $f_8 = 1-f_0 - f_6$,
the condition for WSLS to give the highest expected payoff is summarized as
the intersection of the following two inequalities [Fig.~\ref{fig:simplex}(a)]:
\begin{eqnarray}
f_6 &>& \frac{1}{3}f_8 - \left(\frac{5c}{4+3c}\right) \label{eq:ineq1}\\
f_6 &>& \left( \frac{3c}{2+3c}\right)- \frac{3}{5} \left(\frac{2-c}{2+3c}
\right) f_8 \label{eq:ineq2}.
\end{eqnarray}
The above inequalities are written for $f_6$ because it is $\mathbf{d}_6$ that
has WSLS as the best response (Table~\ref{table:br}).
If $c>1/3$, the former inequality becomes trivial because of the
positivity of $f_6$. Note that WSLS still gives the highest expected payoff for
a significant part of the simplex even when the cost of cooperation is as high
as $c=0.9$ [Fig.~\ref{fig:simplex}(b)].

Similarly, we can check what an observer would conclude after observing
nearly cooperation only,
although it is of less importance compared with the above
case of a defecting population (Fig.~\ref{fig:best16_uncertainty}).
For a general prior represented by $(f_9, f_{14}, f_{15})$, where
$f_{14} = 1-f_9 - f_{15}$, WSLS gives the highest expected payoff when
\begin{equation}
f_9 > \left( \frac{c}{1-c} \right) \left( 1 + \frac{f_{15}}{2} \right),
\end{equation}
as can be seen in Fig.~\ref{fig:simplex}(c). This inequality can be satisfied
only if $c \le 1/2$: Otherwise, it is better off to
be a defector by playing $\mathbf{d}_0$ or $\mathbf{d}_1$.
[Fig.~\ref{fig:simplex}(d)].

\section{Summary and Discussion}

In summary, we have investigated the iterated PD game in terms of best-response
relations and checked how it is modified by observational learning. Thereby we
have addressed a question about how cooperation is affected by cultural
transmission, which may be systematically involved with observational
uncertainty. The notion of SCE takes this systematic uncertainty into account,
and its intersection with NE can be an equilibrium refinement.
It is worth pointing out the following: If everyone plays a certain strategy
$\mathbf{d}_i$ with belief that everyone else does the same, the whole
situation is self-consistent in the sense that observation will always confirm
the belief, which in turn agrees with the actual behaviour. The importance of
SCENE becomes clear when someone happens to play a different strategy or
begins to doubt the belief: If $\mathbf{d}_i$ is not a NE, the player will benefit
from the deviant behaviour and reinforce it. If $\mathbf{d}_i$ is not a SCE, the
player may fail to dispel the doubt, which will undermine the prevailing
culture. Therefore, the strategy has to be a SCENE for being transmitted in a
stable manner through observational learning.

As a reference point, we have started with the conventional assumption that one
can identify a strategy without uncertainty, and checked the best-response
relations within the set of M$_1$ pure strategies.
Our finding is
that a symmetric NE is possible if one uses one of the following three
strategies: AllD, GT$_1$, and WSLS (Fig.~\ref{fig:best16}). Only the last one is
efficient. Although we have restricted ourselves to pure strategies, we can
discuss the idea behind it as follows: Let us consider a monomorphic population
playing a mixed strategy $\mathbf{q} = [q_{_{CC}}, q_{_{CD}}, q_{_{DC}},
q_{_{DD}}]$, where each element means the probability to cooperate in a given
circumstance.
Such a mixed strategy can be represented as a
point inside a four-dimensional unit
hypercube. The observer seeks the best response to it, say, $\mathbf{p} =
[p_{_{CC}}, p_{_{CD}}, p_{_{DC}}, p_{_{DD}}]$. Suppose that $\mathbf{p}$ also
turns out to be a mixed strategy, say, containing $\mathbf{d}_k$ and $\mathbf{d}_l$
with $k \neq l$. According to the Bishop-Cannings
theorem~\cite{bishop1978generalized}, it implies that
\begin{equation}
\Pi(\mathbf{d}_k, \mathbf{q}, \epsilon) = \Pi(\mathbf{d}_l, \mathbf{q}, \epsilon),
\label{eq:bishop}
\end{equation}
and this equality imposes a set of constraints on $\mathbf{q}$, rendering the
dimensionality of the solution manifold lower than four. Therefore, to almost
all $\mathbf{q}$ in the four-dimensional hypercube, only one pure strategy will be
found as the best response. In Appendix~\ref{app:reactive}, we provide an
explicit proof for this argument in case of reactive strategies.

Even if our theoretical framework of Bayesian best-response
dynamics is an idealization, we believe that it captures certain aspects of
animal behaviour. For example, although the best-response dynamics {\it per se}
shows poor performance in explaining learning
behaviour because of its deterministic character~\cite{nagel1998experimental},
its modified versions can provide reasonable description
for experimental results~\cite{van1997origin,cheung1997individual}.
In addition, some studies show that
Bayesian updating yields consistent results with observed behaviour of
animals, including mammals, birds, a fish and an insect, in the foraging
and reproduction activities~\cite{valone2006animals}.
These studies support the
Bayesian brain hypothesis, which argues that the brain has to successfully
simulate the external world in which Bayes' theorem
holds~\cite{friston2012history}.
We also point out that the posterior can be calculated correctly even if the
observer has short-term memory as implied by the M$_1$ assumption:
As long as input observations are exchangeable with each other, Bayesian
updating can be done in a sequential manner, i.e., by modifying the prior little
by little every time a new observation arrives, and it is mathematically
equivalent to a batch update that uses all the observations at once.

To conclude,
if we take observational learning into consideration,
our result suggests that WSLS can be a SCENE to a Bayesian observer, whereas
AllD cannot under observational uncertainty.
That is, if the number of observations is too small
to see how to behave after error, the uncertainty provides a way to escape
from full defection, whereas WSLS can still maintain cooperation:
The point is that AllD is not easy to learn by observing defectors
because it is difficult to tell what they would choose if someone
actually cooperated. WSLS is also
difficult to learn, but the uncertainty works in an asymmetric way
because one can expect more from mutual
cooperation than from full defection by the very definition of
the PD game.




\appendix
\section{Stationary distribution}
\label{app:stationary}
Let us consider two players, Alice and Bob, playing the PD game repeatedly.
As written in
Eq.~\eqref{eq:effective}, Alice's effective behaviour in the noisy environment
is described by a mixed strategy
$\mathbf{s}_A^\epsilon = (q_{_{CC}}, q_{_{CD}}, q_{_{DC}}, q_{_{DD}})$,
where $q_{_{XY}} \in \{\epsilon, 1-\epsilon\}$ denotes Alice's probability of
cooperation when she and Bob did $X$ and $Y$, respectively, in the previous
round. In the same manner, another mixed strategy
$\mathbf{s}_B^\epsilon = (r_{_{CC}}, r_{_{CD}}, r_{_{DC}}, r_{_{DD}})$ applies
to Bob's effective behaviour [Eq.~\eqref{eq:effective2}], where $r_{_{XY}} \in
\{\epsilon, 1-\epsilon\}$ denotes Bob's probability of cooperation when he and
Alice did $X$ and $Y$, respectively, in the previous round.
Let $v^{(t)}_{_{XY}}$ be the probability to see Alice and Bob choosing $X$
and $Y$, respectively, in round $t$.
The condition $v^{(t)}_{_{CC}}+ v^{(t)}_{_{CD}}+
v^{(t)}_{_{DC}}+ v^{(t)}_{_{DD}}=1$ is satisfied all the time.
The probability distribution
$\mathbf{v}^{(t)} \equiv \left(v^{(t)}_{_{CC}}, v^{(t)}_{_{CD}},
v^{(t)}_{_{DC}}, v^{(t)}_{_{DD}}\right)$ evolves as $\mathbf{v}^{(t+1)} = W
\mathbf{v}^{(t)}$ with
\begin{equation}
W =
\begin{bmatrix}
q_{_{CC}} r_{_{CC}} &
q_{_{CD}} r_{_{DC}} &
q_{_{DC}} r_{_{CD}} &
q_{_{DD}} r_{_{DD}} \\
q_{_{CC}} \bar{r}_{_{CC}} &
q_{_{CD}} \bar{r}_{_{DC}} &
q_{_{DC}} \bar{r}_{_{CD}} &
q_{_{DD}} \bar{r}_{_{DD}} \\
\bar{q}_{_{CC}} r_{_{CC}} &
\bar{q}_{_{CD}} r_{_{DC}} &
\bar{q}_{_{DC}} r_{_{CD}} &
\bar{q}_{_{DD}} r_{_{DD}} \\
\bar{q}_{_{CC}} \bar{r}_{_{CC}} &
\bar{q}_{_{CD}} \bar{r}_{_{DC}} &
\bar{q}_{_{DC}} \bar{r}_{_{CD}} &
\bar{q}_{_{DD}} \bar{r}_{_{DD}}
\end{bmatrix},
\end{equation}
where $\bar{q}_{_{XY}} \equiv 1-q_{_{XY}}$ and $\bar{r}_{_{XY}} \equiv
1-r_{_{XY}}$. Note that it is a positive stochastic matrix for $\epsilon>0$.
According to the Perron-Frobenius theorem, it has a unique largest eigenvalue
$1$, and the corresponding eigenvector can be chosen to have positive entries.
Thus, by solving $W\mathbf{v} = \mathbf{v}$, we can obtain the stationary
distribution $\mathbf{v}=\left(v_{_{CC}}, v_{_{CD}}, v_{_{DC}},
v_{_{DD}}\right)$.
Each element $v_{_{XY}}$ can be interpreted as the
long-time average frequency of $XY$, and
it can readily be expanded as a Taylor series in terms of $\epsilon$. To
determine the best response to $\mathbf{d}_\beta$ as shown in
Table~\ref{table:br}, we calculate the long-term
average payoff of $\mathbf{d}_\alpha$ against it for every $\alpha \in
\{0, \ldots, 15\}$ [Eq.~\eqref{eq:avgpay}] and compare the Taylor-expanded
expressions order by order. As for Table~\ref{table:vXY}, we set $\alpha =
\beta$ and retain only the leading order terms in the Taylor series for
$\mathbf{v}$.

\section{Bayesian inference}
\label{app:bayes}
To illustrate the inference procedure, let us
assume that $\mathbf{v} \approx (0,0,0,1)$ is given to Alice. She has a set of
candidate strategies $\Lambda \equiv \{ \mathbf{d}_{0}, \mathbf{d}_{6}, \mathbf{d}_{8}
\}$ for the resident strategy $\mathbf{q}$.
Alice assigns equal prior probability to each of these
candidate strategies. In a certain round $t$, she observes interaction between
Eve and Frank both of whom use $\mathbf{q}$. Let $E_t$ and $F_t$ denote Eve's
and Frank's moves, respectively, in round $t$. If Alice sees Eve cooperate,
i.e., $E_t = C$,
after $S_{t-1} \equiv (E_{t-1}, F_{t-1}) = (C,C)$, she may use this additional
information in a Bayesian way to calculate the posterior probability of
$\mathbf{q} = \mathbf{d}_{0}$ as follows:
\begin{eqnarray}
&&P \left( \mathbf{q} = \mathbf{d}_{0} | E_t, S_{t-1} \right)
= \frac
{P \left( E_t | S_{t-1}, \mathbf{d}_{0} \right) P \left( S_{t-1} |
\mathbf{d}_{0} \right) P\left( \mathbf{d}_{0} \right) }
{\sum_{\mathbf{d}_{i} \in \Lambda} P \left( E_t | S_{t-1}, \mathbf{d}_{i} \right) P
\left( S_{t-1} | \mathbf{d}_{i} \right) P\left( \mathbf{d}_{i} \right) }\\
&&= \frac{\epsilon \cdot \epsilon^2 \cdot (1/3)}{\epsilon \cdot \epsilon^2
\cdot (1/3) + \epsilon \cdot (2\epsilon-5\epsilon^2+4\epsilon^3) \cdot (1/3)
+ \epsilon \cdot \epsilon/2 \cdot (1/3)},
\end{eqnarray}
where $P(E_t | S_{t-1}, \mathbf{d}_{i})$ is directly obtained from $\mathbf{d}_{i}$,
and $P(S_{t-1} | \mathbf{d}_{i})$ is taken from the stationary probability
distribution $\mathbf{v}$.
This posterior probability is used as prior probability for the next
observation. If $\mathbf{q}$ is actually $\mathbf{d}_{6}$, the average
number of times to observe $E_t=C$ after $S_{t-1}=(C,C)$ will be
\begin{equation}
M P(E_t, S_{t-1} | \mathbf{q} = \mathbf{d}_{6}) =
M P(E_t | S_{t-1}, \mathbf{d}_{6}) P(S_{t-1}| \mathbf{d}_{6}).
\end{equation}
In this way, Alice obtains the final posterior probability of $\mathbf{q} =
\mathbf{d}_{0}$ after observing interaction between $M$ pairs of players,
when their actual strategy is $\mathbf{d}_{6}$.
If $\epsilon$ is fixed as a small positive value, this inference procedure
approaches the correct answer as $M \to \infty$.
The effect of observational uncertainty manifests itself when $M \epsilon \ll
1$. For example, we may choose $M \approx \epsilon^{-1/2}$ as a representative
value for $1 \ll M \ll \epsilon^{-1}$ and check various values of $\epsilon$
from $10^{-2}$ to $10^{-6}$. Then, the above calculation confirms that the
posterior probabilities should remain identical to the prior ones due to the
lack of observation.

\section{Best-response relations among reactive strategies}
\label{app:reactive}
Let us consider two
reactive strategies $\mathbf{p} = [p_{_C}, p_{_D}, p_{_C}, p_{_D}]$
and $\mathbf{q} = [q_{_C}, q_{_D}, q_{_C}, q_{_D}]$. The long-term average payoff
of $\mathbf{p}$ against $\mathbf{q}$ is
\begin{equation}
\Pi  = \frac{(p_{_D} q_{_C} - p_{_D} q_{_D} + q_{_D}) - c(p_{_D} + p_{_C} q_{_D}
- p_{_D} q_{_D})}{1 - (p_{_C} - p_{_D}) (q_{_C} - q_{_D})}
\end{equation}
in the limit of $\epsilon \to 0$.
After some algebra, we find the following: First,
if $q_{_{C}} - q_{_{D}} > c$,
both $\partial \Pi / \partial p_{_{C}}$ and
$\partial \Pi / \partial p_{_{D}}$ are positive, so the best response
is given by $p_{_C} = p_{_D} = 1$.
Or, if $q_{_{C}} - q_{_{D}} < c$, both
$\partial \Pi / \partial p_{_{C}}$ and
$\partial \Pi / \partial p_{_{D}}$ are negative, so the best response
is given by $p_{_C} = p_{_D} = 0$.
Note that we have neglected the measure-zero line defined by $q_{_{C}} -
q_{_{D}} = c$, on which the best response is not uniquely determined.




\begin{thebibliography}{27}%
\makeatletter
\providecommand \@ifxundefined [1]{%
 \@ifx{#1\undefined}
}%
\providecommand \@ifnum [1]{%
 \ifnum #1\expandafter \@firstoftwo
 \else \expandafter \@secondoftwo
 \fi
}%
\providecommand \@ifx [1]{%
 \ifx #1\expandafter \@firstoftwo
 \else \expandafter \@secondoftwo
 \fi
}%
\providecommand \natexlab [1]{#1}%
\providecommand \enquote  [1]{``#1''}%
\providecommand \bibnamefont  [1]{#1}%
\providecommand \bibfnamefont [1]{#1}%
\providecommand \citenamefont [1]{#1}%
\providecommand \href@noop [0]{\@secondoftwo}%
\providecommand \href [0]{\begingroup \@sanitize@url \@href}%
\providecommand \@href[1]{\@@startlink{#1}\@@href}%
\providecommand \@@href[1]{\endgroup#1\@@endlink}%
\providecommand \@sanitize@url [0]{\catcode `\\12\catcode `\$12\catcode
  `\&12\catcode `\#12\catcode `\^12\catcode `\_12\catcode `\%12\relax}%
\providecommand \@@startlink[1]{}%
\providecommand \@@endlink[0]{}%
\providecommand \url  [0]{\begingroup\@sanitize@url \@url }%
\providecommand \@url [1]{\endgroup\@href {#1}{\urlprefix }}%
\providecommand \urlprefix  [0]{URL }%
\providecommand \Eprint [0]{\href }%
\providecommand \doibase [0]{http://dx.doi.org/}%
\providecommand \selectlanguage [0]{\@gobble}%
\providecommand \bibinfo  [0]{\@secondoftwo}%
\providecommand \bibfield  [0]{\@secondoftwo}%
\providecommand \translation [1]{[#1]}%
\providecommand \BibitemOpen [0]{}%
\providecommand \bibitemStop [0]{}%
\providecommand \bibitemNoStop [0]{.\EOS\space}%
\providecommand \EOS [0]{\spacefactor3000\relax}%
\providecommand \BibitemShut  [1]{\csname bibitem#1\endcsname}%
\let\auto@bib@innerbib\@empty
\bibitem [{\citenamefont {{Maynard~Smith}}(1982)}]{smith1982evolution}%
  \BibitemOpen
  \bibfield  {author} {\bibinfo {author} {\bibfnamefont {J.}~\bibnamefont
  {{Maynard~Smith}}},\ }\href@noop {} {\emph {\bibinfo {title} {Evolution and
  the Theory of Games}}}\ (\bibinfo  {publisher} {Cambridge University Press},\
  \bibinfo {address} {Cambridge, UK},\ \bibinfo {year} {1982})\BibitemShut
  {NoStop}%
\bibitem [{\citenamefont {Kasper}\ \emph {et~al.}(2017)\citenamefont {Kasper},
  \citenamefont {Vierbuchen}, \citenamefont {Ernst}, \citenamefont {Fischer},
  \citenamefont {Radersma}, \citenamefont {Raulo}, \citenamefont
  {Cunha-Saraiva}, \citenamefont {Wu}, \citenamefont {Mobley},\ and\
  \citenamefont {Taborsky}}]{kasper2017genetics}%
  \BibitemOpen
  \bibfield  {author} {\bibinfo {author} {\bibfnamefont {C.}~\bibnamefont
  {Kasper}}, \bibinfo {author} {\bibfnamefont {M.}~\bibnamefont {Vierbuchen}},
  \bibinfo {author} {\bibfnamefont {U.}~\bibnamefont {Ernst}}, \bibinfo
  {author} {\bibfnamefont {S.}~\bibnamefont {Fischer}}, \bibinfo {author}
  {\bibfnamefont {R.}~\bibnamefont {Radersma}}, \bibinfo {author}
  {\bibfnamefont {A.}~\bibnamefont {Raulo}}, \bibinfo {author} {\bibfnamefont
  {F.}~\bibnamefont {Cunha-Saraiva}}, \bibinfo {author} {\bibfnamefont
  {M.}~\bibnamefont {Wu}}, \bibinfo {author} {\bibfnamefont {K.~B.}\
  \bibnamefont {Mobley}}, \ and\ \bibinfo {author} {\bibfnamefont
  {B.}~\bibnamefont {Taborsky}},\ }\href@noop {} {\bibfield  {journal}
  {\bibinfo  {journal} {Mol. Ecol.}\ }\textbf {\bibinfo {volume} {26}},\
  \bibinfo {pages} {4364} (\bibinfo {year} {2017})}\BibitemShut {NoStop}%
\bibitem [{\citenamefont {Manfredini}\ \emph {et~al.}(2018)\citenamefont
  {Manfredini}, \citenamefont {Brown},\ and\ \citenamefont
  {Toth}}]{manfredini2018candidate}%
  \BibitemOpen
  \bibfield  {author} {\bibinfo {author} {\bibfnamefont {F.}~\bibnamefont
  {Manfredini}}, \bibinfo {author} {\bibfnamefont {M.~J.}\ \bibnamefont
  {Brown}}, \ and\ \bibinfo {author} {\bibfnamefont {A.~L.}\ \bibnamefont
  {Toth}},\ }\href@noop {} {\bibfield  {journal} {\bibinfo  {journal} {J. Comp.
  Physiol. A}\ }\textbf {\bibinfo {volume} {204}},\ \bibinfo {pages} {449}
  (\bibinfo {year} {2018})}\BibitemShut {NoStop}%
\bibitem [{\citenamefont {Bandura}(1977)}]{bandura1977social}%
  \BibitemOpen
  \bibfield  {author} {\bibinfo {author} {\bibfnamefont {A.}~\bibnamefont
  {Bandura}},\ }\href@noop {} {\emph {\bibinfo {title} {Social Learning
  Theory}}}\ (\bibinfo  {publisher} {Prentice Hall},\ \bibinfo {address}
  {Englewood Cliffs, NJ},\ \bibinfo {year} {1977})\BibitemShut {NoStop}%
\bibitem [{\citenamefont {Kr{\"u}tzen}\ \emph {et~al.}(2005)\citenamefont
  {Kr{\"u}tzen}, \citenamefont {Mann}, \citenamefont {Heithaus}, \citenamefont
  {Connor}, \citenamefont {Bejder},\ and\ \citenamefont
  {Sherwin}}]{krutzen2005cultural}%
  \BibitemOpen
  \bibfield  {author} {\bibinfo {author} {\bibfnamefont {M.}~\bibnamefont
  {Kr{\"u}tzen}}, \bibinfo {author} {\bibfnamefont {J.}~\bibnamefont {Mann}},
  \bibinfo {author} {\bibfnamefont {M.~R.}\ \bibnamefont {Heithaus}}, \bibinfo
  {author} {\bibfnamefont {R.~C.}\ \bibnamefont {Connor}}, \bibinfo {author}
  {\bibfnamefont {L.}~\bibnamefont {Bejder}}, \ and\ \bibinfo {author}
  {\bibfnamefont {W.~B.}\ \bibnamefont {Sherwin}},\ }\href@noop {} {\bibfield
  {journal} {\bibinfo  {journal} {Proc. Natl. Acad. Sci. USA}\ }\textbf
  {\bibinfo {volume} {102}},\ \bibinfo {pages} {8939} (\bibinfo {year}
  {2005})}\BibitemShut {NoStop}%
\bibitem [{\citenamefont {Frith}\ and\ \citenamefont
  {Frith}(2012)}]{frith2012mechanisms}%
  \BibitemOpen
  \bibfield  {author} {\bibinfo {author} {\bibfnamefont {C.~D.}\ \bibnamefont
  {Frith}}\ and\ \bibinfo {author} {\bibfnamefont {U.}~\bibnamefont {Frith}},\
  }\href@noop {} {\bibfield  {journal} {\bibinfo  {journal} {Annu. Rev.
  Psychol.}\ }\textbf {\bibinfo {volume} {63}},\ \bibinfo {pages} {287}
  (\bibinfo {year} {2012})}\BibitemShut {NoStop}%
\bibitem [{\citenamefont {Di~Pellegrino}\ \emph {et~al.}(1992)\citenamefont
  {Di~Pellegrino}, \citenamefont {Fadiga}, \citenamefont {Fogassi},
  \citenamefont {Gallese},\ and\ \citenamefont
  {Rizzolatti}}]{di1992understanding}%
  \BibitemOpen
  \bibfield  {author} {\bibinfo {author} {\bibfnamefont {G.}~\bibnamefont
  {Di~Pellegrino}}, \bibinfo {author} {\bibfnamefont {L.}~\bibnamefont
  {Fadiga}}, \bibinfo {author} {\bibfnamefont {L.}~\bibnamefont {Fogassi}},
  \bibinfo {author} {\bibfnamefont {V.}~\bibnamefont {Gallese}}, \ and\
  \bibinfo {author} {\bibfnamefont {G.}~\bibnamefont {Rizzolatti}},\
  }\href@noop {} {\bibfield  {journal} {\bibinfo  {journal} {Exp. Brain Res.}\
  }\textbf {\bibinfo {volume} {91}},\ \bibinfo {pages} {176} (\bibinfo {year}
  {1992})}\BibitemShut {NoStop}%
\bibitem [{\citenamefont {Gallese}\ \emph {et~al.}(1996)\citenamefont
  {Gallese}, \citenamefont {Fadiga}, \citenamefont {Fogassi},\ and\
  \citenamefont {Rizzolatti}}]{gallese1996action}%
  \BibitemOpen
  \bibfield  {author} {\bibinfo {author} {\bibfnamefont {V.}~\bibnamefont
  {Gallese}}, \bibinfo {author} {\bibfnamefont {L.}~\bibnamefont {Fadiga}},
  \bibinfo {author} {\bibfnamefont {L.}~\bibnamefont {Fogassi}}, \ and\
  \bibinfo {author} {\bibfnamefont {G.}~\bibnamefont {Rizzolatti}},\
  }\href@noop {} {\bibfield  {journal} {\bibinfo  {journal} {Brain}\ }\textbf
  {\bibinfo {volume} {119}},\ \bibinfo {pages} {593} (\bibinfo {year}
  {1996})}\BibitemShut {NoStop}%
\bibitem [{\citenamefont {Ferrari}\ and\ \citenamefont
  {Rizzolatti}(2014)}]{ferrari2014mirror}%
  \BibitemOpen
  \bibfield  {author} {\bibinfo {author} {\bibfnamefont {P.~F.}\ \bibnamefont
  {Ferrari}}\ and\ \bibinfo {author} {\bibfnamefont {G.}~\bibnamefont
  {Rizzolatti}},\ }\href@noop {} {\bibfield  {journal} {\bibinfo  {journal}
  {Philos. Trans. R. Soc. Lond. B}\ }\textbf {\bibinfo {volume} {369}},\
  \bibinfo {pages} {20130169} (\bibinfo {year} {2014})}\BibitemShut {NoStop}%
\bibitem [{\citenamefont {Leimar}\ and\ \citenamefont
  {McNamara}(2015)}]{leimar2015evolution}%
  \BibitemOpen
  \bibfield  {author} {\bibinfo {author} {\bibfnamefont {O.}~\bibnamefont
  {Leimar}}\ and\ \bibinfo {author} {\bibfnamefont {J.~M.}\ \bibnamefont
  {McNamara}},\ }\href@noop {} {\bibfield  {journal} {\bibinfo  {journal} {Am.
  Nat.}\ }\textbf {\bibinfo {volume} {185}},\ \bibinfo {pages} {E55} (\bibinfo
  {year} {2015})}\BibitemShut {NoStop}%
\bibitem [{\citenamefont {Fudenberg}\ and\ \citenamefont
  {Levine}(1998)}]{fudenberg1998theory}%
  \BibitemOpen
  \bibfield  {author} {\bibinfo {author} {\bibfnamefont {D.}~\bibnamefont
  {Fudenberg}}\ and\ \bibinfo {author} {\bibfnamefont {D.~K.}\ \bibnamefont
  {Levine}},\ }\href@noop {} {\emph {\bibinfo {title} {The Theory of Learning
  in Games}}}\ (\bibinfo  {publisher} {MIT Press},\ \bibinfo {address}
  {Cambridge, MA},\ \bibinfo {year} {1998})\BibitemShut {NoStop}%
\bibitem [{\citenamefont {Sandholm}(2001)}]{sandholm2001almost}%
  \BibitemOpen
  \bibfield  {author} {\bibinfo {author} {\bibfnamefont {W.~H.}\ \bibnamefont
  {Sandholm}},\ }\href@noop {} {\bibfield  {journal} {\bibinfo  {journal} {Int.
  J. Game Theory}\ }\textbf {\bibinfo {volume} {30}},\ \bibinfo {pages} {107}
  (\bibinfo {year} {2001})}\BibitemShut {NoStop}%
\bibitem [{\citenamefont {Kreindler}\ and\ \citenamefont
  {Young}(2013)}]{kreindler2013fast}%
  \BibitemOpen
  \bibfield  {author} {\bibinfo {author} {\bibfnamefont {G.~E.}\ \bibnamefont
  {Kreindler}}\ and\ \bibinfo {author} {\bibfnamefont {H.~P.}\ \bibnamefont
  {Young}},\ }\href@noop {} {\bibfield  {journal} {\bibinfo  {journal} {Games
  Econ. Behav.}\ }\textbf {\bibinfo {volume} {80}},\ \bibinfo {pages} {39}
  (\bibinfo {year} {2013})}\BibitemShut {NoStop}%
\bibitem [{\citenamefont {Baek}\ \emph {et~al.}(2016)\citenamefont {Baek},
  \citenamefont {Jeong}, \citenamefont {Hilbe},\ and\ \citenamefont
  {Nowak}}]{baek2016comparing}%
  \BibitemOpen
  \bibfield  {author} {\bibinfo {author} {\bibfnamefont {S.~K.}\ \bibnamefont
  {Baek}}, \bibinfo {author} {\bibfnamefont {H.-C.}\ \bibnamefont {Jeong}},
  \bibinfo {author} {\bibfnamefont {C.}~\bibnamefont {Hilbe}}, \ and\ \bibinfo
  {author} {\bibfnamefont {M.~A.}\ \bibnamefont {Nowak}},\ }\href@noop {}
  {\bibfield  {journal} {\bibinfo  {journal} {Sci. Rep.}\ }\textbf {\bibinfo
  {volume} {6}},\ \bibinfo {pages} {25676} (\bibinfo {year}
  {2016})}\BibitemShut {NoStop}%
\bibitem [{\citenamefont {Kraines}\ and\ \citenamefont
  {Kraines}(1989)}]{kraines1989pavlov}%
  \BibitemOpen
  \bibfield  {author} {\bibinfo {author} {\bibfnamefont {D.}~\bibnamefont
  {Kraines}}\ and\ \bibinfo {author} {\bibfnamefont {V.}~\bibnamefont
  {Kraines}},\ }\href@noop {} {\bibfield  {journal} {\bibinfo  {journal}
  {Theory Decis.}\ }\textbf {\bibinfo {volume} {26}},\ \bibinfo {pages} {47}
  (\bibinfo {year} {1989})}\BibitemShut {NoStop}%
\bibitem [{\citenamefont {Nowak}\ and\ \citenamefont
  {Sigmund}(1993)}]{nowak1993strategy}%
  \BibitemOpen
  \bibfield  {author} {\bibinfo {author} {\bibfnamefont {M.}~\bibnamefont
  {Nowak}}\ and\ \bibinfo {author} {\bibfnamefont {K.}~\bibnamefont
  {Sigmund}},\ }\href@noop {} {\bibfield  {journal} {\bibinfo  {journal}
  {Nature}\ }\textbf {\bibinfo {volume} {364}},\ \bibinfo {pages} {56}
  (\bibinfo {year} {1993})}\BibitemShut {NoStop}%
\bibitem [{\citenamefont {Imhof}\ \emph {et~al.}(2007)\citenamefont {Imhof},
  \citenamefont {Fudenberg},\ and\ \citenamefont {Nowak}}]{imhof2007tit}%
  \BibitemOpen
  \bibfield  {author} {\bibinfo {author} {\bibfnamefont {L.~A.}\ \bibnamefont
  {Imhof}}, \bibinfo {author} {\bibfnamefont {D.}~\bibnamefont {Fudenberg}}, \
  and\ \bibinfo {author} {\bibfnamefont {M.~A.}\ \bibnamefont {Nowak}},\
  }\href@noop {} {\bibfield  {journal} {\bibinfo  {journal} {J. Theor. Biol.}\
  }\textbf {\bibinfo {volume} {247}},\ \bibinfo {pages} {574} (\bibinfo {year}
  {2007})}\BibitemShut {NoStop}%
\bibitem [{\citenamefont {Nowak}(1990)}]{nowak1990stochastic}%
  \BibitemOpen
  \bibfield  {author} {\bibinfo {author} {\bibfnamefont {M.}~\bibnamefont
  {Nowak}},\ }\href@noop {} {\bibfield  {journal} {\bibinfo  {journal} {Theor.
  Popul. Biol.}\ }\textbf {\bibinfo {volume} {38}},\ \bibinfo {pages} {93}
  (\bibinfo {year} {1990})}\BibitemShut {NoStop}%
\bibitem [{\citenamefont {Nowak}\ \emph {et~al.}(1995)\citenamefont {Nowak},
  \citenamefont {Sigmund},\ and\ \citenamefont {El-Sedy}}]{nowak1995automata}%
  \BibitemOpen
  \bibfield  {author} {\bibinfo {author} {\bibfnamefont {M.~A.}\ \bibnamefont
  {Nowak}}, \bibinfo {author} {\bibfnamefont {K.}~\bibnamefont {Sigmund}}, \
  and\ \bibinfo {author} {\bibfnamefont {E.}~\bibnamefont {El-Sedy}},\
  }\href@noop {} {\bibfield  {journal} {\bibinfo  {journal} {J. Math. Biol.}\
  }\textbf {\bibinfo {volume} {33}},\ \bibinfo {pages} {703} (\bibinfo {year}
  {1995})}\BibitemShut {NoStop}%
\bibitem [{\citenamefont {Press}\ and\ \citenamefont
  {Dyson}(2012)}]{press2012iterated}%
  \BibitemOpen
  \bibfield  {author} {\bibinfo {author} {\bibfnamefont {W.~H.}\ \bibnamefont
  {Press}}\ and\ \bibinfo {author} {\bibfnamefont {F.~J.}\ \bibnamefont
  {Dyson}},\ }\href@noop {} {\bibfield  {journal} {\bibinfo  {journal} {Proc.
  Natl. Acad. Sci. USA}\ }\textbf {\bibinfo {volume} {109}},\ \bibinfo {pages}
  {10409} (\bibinfo {year} {2012})}\BibitemShut {NoStop}%
\bibitem [{\citenamefont {Harper}\ \emph {et~al.}(2015)\citenamefont {Harper}
  \emph {et~al.}}]{pythonternary}%
  \BibitemOpen
  \bibfield  {author} {\bibinfo {author} {\bibfnamefont {M.}~\bibnamefont
  {Harper}} \emph {et~al.},\ }\href@noop {} {\enquote {\bibinfo {title}
  {python-ternary: Ternary plots in python},}\ } (\bibinfo {year} {2015}),\
  \bibinfo {note} {zenodo 10.5281/zenodo.594435}\BibitemShut {NoStop}%
\bibitem [{\citenamefont {Bishop}\ and\ \citenamefont
  {Cannings}(1978)}]{bishop1978generalized}%
  \BibitemOpen
  \bibfield  {author} {\bibinfo {author} {\bibfnamefont {D.}~\bibnamefont
  {Bishop}}\ and\ \bibinfo {author} {\bibfnamefont {C.}~\bibnamefont
  {Cannings}},\ }\href@noop {} {\bibfield  {journal} {\bibinfo  {journal} {J.
  Theor. Biol.}\ }\textbf {\bibinfo {volume} {70}},\ \bibinfo {pages} {85}
  (\bibinfo {year} {1978})}\BibitemShut {NoStop}%
\bibitem [{\citenamefont {Nagel}\ and\ \citenamefont
  {Tang}(1998)}]{nagel1998experimental}%
  \BibitemOpen
  \bibfield  {author} {\bibinfo {author} {\bibfnamefont {R.}~\bibnamefont
  {Nagel}}\ and\ \bibinfo {author} {\bibfnamefont {F.~F.}\ \bibnamefont
  {Tang}},\ }\href@noop {} {\bibfield  {journal} {\bibinfo  {journal} {J. Math.
  Psychol.}\ }\textbf {\bibinfo {volume} {42}},\ \bibinfo {pages} {356}
  (\bibinfo {year} {1998})}\BibitemShut {NoStop}%
\bibitem [{\citenamefont {Van~Huyck}\ \emph {et~al.}(1997)\citenamefont
  {Van~Huyck}, \citenamefont {Battalio},\ and\ \citenamefont
  {Rankin}}]{van1997origin}%
  \BibitemOpen
  \bibfield  {author} {\bibinfo {author} {\bibfnamefont {J.~B.}\ \bibnamefont
  {Van~Huyck}}, \bibinfo {author} {\bibfnamefont {R.~C.}\ \bibnamefont
  {Battalio}}, \ and\ \bibinfo {author} {\bibfnamefont {F.~W.}\ \bibnamefont
  {Rankin}},\ }\href@noop {} {\bibfield  {journal} {\bibinfo  {journal} {Econ.
  J.}\ }\textbf {\bibinfo {volume} {107}},\ \bibinfo {pages} {576} (\bibinfo
  {year} {1997})}\BibitemShut {NoStop}%
\bibitem [{\citenamefont {Cheung}\ and\ \citenamefont
  {Friedman}(1997)}]{cheung1997individual}%
  \BibitemOpen
  \bibfield  {author} {\bibinfo {author} {\bibfnamefont {Y.-W.}\ \bibnamefont
  {Cheung}}\ and\ \bibinfo {author} {\bibfnamefont {D.}~\bibnamefont
  {Friedman}},\ }\href@noop {} {\bibfield  {journal} {\bibinfo  {journal}
  {Games Econ. Behav.}\ }\textbf {\bibinfo {volume} {19}},\ \bibinfo {pages}
  {46} (\bibinfo {year} {1997})}\BibitemShut {NoStop}%
\bibitem [{\citenamefont {J.~Valone}(2006)}]{valone2006animals}%
  \BibitemOpen
  \bibfield  {author} {\bibinfo {author} {\bibfnamefont {T.}~\bibnamefont
  {J.~Valone}},\ }\href@noop {} {\bibfield  {journal} {\bibinfo  {journal}
  {Oikos}\ }\textbf {\bibinfo {volume} {112}},\ \bibinfo {pages} {252}
  (\bibinfo {year} {2006})}\BibitemShut {NoStop}%
\bibitem [{\citenamefont {Friston}(2012)}]{friston2012history}%
  \BibitemOpen
  \bibfield  {author} {\bibinfo {author} {\bibfnamefont {K.}~\bibnamefont
  {Friston}},\ }\href@noop {} {\bibfield  {journal} {\bibinfo  {journal}
  {NeuroImage}\ }\textbf {\bibinfo {volume} {62}},\ \bibinfo {pages} {1230}
  (\bibinfo {year} {2012})}\BibitemShut {NoStop}%
\end{thebibliography}
%
\end{document}